# Modelling the transient processes produced under heavy particle irradiation

Sorina Lazanu [a1], Ionel Lazanu [b], Gheorghe Ciobanu [b]

[a] *National Institute of Materials Physics, P.O. Box MG-7, Bucharest-Măgurele, Romania*
[b] *University of Bucharest, Faculty of Physics, P.O. Box MG-11, Bucharest-Măgurele, Romania*

**Abstract**

A new model for the thermal spike produced by the nuclear energy loss, as source of transient processes, is derived analytically, for power law dependences of the diffusivity on temperature, as solution of the heat equation. The contribution of the ionizing energy loss to the spike is not included. The range of validity of the model is analysed, and the results are compared with numerical solutions obtained in the frame of the previous model of the authors, which takes into account both nuclear and ionization energy losses, as well as the coupling between the two subsystems in crystalline semiconductors. Particular solutions are discussed and the errors induced by these approximations are analysed.

**Keywords**: Energy loss; transient processes; thermal spike model; ions; semiconductors; astroparticle applications.

## 1. Introduction

The theoretical aspects of the penetration of atomic particles through matter were established long ago by Bohr [1], in a distinct formalism by Bethe [2], and have been developed by Lindhard [3], Sigmund [4], Ziegler [5] and others [6 – 10]. Despite this progress, there are still important features concerning the mechanisms related to the energy loss of charged and/or neutral particles in matter which are poorly understood, fact suggested also by a recent book in the field [11].

The absorption of radiation in media is not a continuous but a quantized process obeying statistical laws. The process is transitory, and, as a result, localized regions of the target medium are locally heated and this state evolves due to diffusion mechanisms. The rise in temperature could cause changes in target materials.

---

[1] Address : 105bis Atomistilor Str, Magurele, 77125 Romania ; e-mail: lazanu@infim.ro; phone: +40-213690170; fax: +40-213690177

The analysis of these transient processes and subsequent modifications induced by the local increase of the temperature are interesting both as fundamental physics aspects as well as from the point of view of applied research, because they could be used in new technologies, methods for producing new materials, or in medical applications of radiations.

The consequence of ion irradiation of crystals is the production of lattice disorder and electronic excitation. In the first case, lattice atoms displayed by incident particles could create a cascade of atomic collisions. In spreading the primary recoil energy over a small region of crystal, the cascade presumably produces a local heating effect. Such regions have been considered initially by Seitz [12, 13], and by Brinkman [14], who named them 'thermal spikes' or 'displacement spikes' according to whether or not displaced atoms accompanied the heating effect. Seitz and Koehler [13] defined the thermal spike as being characterized by the fact that the energy transmitted to the lattice by an incident particle could be found in the form of lattice vibrations in such a concentrated way that the local temperatures would be sufficiently high to induce permanent rearrangements of the atoms in the solid up to amorphization. The other extremum, the high electronic energy loss regime, has been treated in the frame of the electronic thermal spike model, with the goal to explain the structural modifications of the material in the track of the ion. The phenomena were successively investigated in metals [15], in insulators [16] and in semiconductors [17]. The thermal spike concept was also considered to explain the stress evolution following ion impact by Hirsch et al. [18] and Müller [19] in ion assisted thin film formation. Although the thermal spike concept treats the solid as a continuum, it provides an estimate of the time scales, critical dimensions, relevant energy densities (i.e. temperatures), and number of thermally activated processes.

In the present paper, an analytical solution for the temperature dependence on time and spatial coordinate during the transient processes produced in a thermal spike is developed. Only one heat source is considered, corresponding to the nuclear energy loss. The solutions for cylindrical and spherical symmetry are obtained by solving the differential equation for the atomic temperature. The model does not consider the electronic contribution to the spike, electron phonon coupling, or other interactions. The range of validity of this approximation is analysed, and some applications of the model for different materials and physical situations for which its application is suitable are presented.

## 2. Mechanisms of energy loss

The processes by which projectiles lose their energy in the medium are discrete due to the quantum character of the physics at this scale. The incoming particle interacts either with the electrons or with the ions, transferring energy to these subsystems, which do also interact. The variation of the energy loss as a function of incoming's particle velocity is generally divided into three regions. At low velocities the main mechanism of energy loss is particle's (elastic) scattering on the atoms placed in their sites in the lattice. Theories based on the statistical model of the atom and on the electron-gas model of metals predict a velocity-proportional behaviour of the rate of electronic energy loss for this region, in agreement with experiments within 20 % accuracy [20, 21]. At high velocities, the projectile loses energy mainly by the ionization of individual target atoms, and Bohr and Bethe-Bloch theories give a good description of the processes. The intermediate energy region could be characterized by the presence of both mechanisms, but a complete theory is not exists yet.

As an illustration, we present in Figure 1 the dependence of the nuclear and electronic energy loss of C, Ge and Pb ions in Si on projectile's kinetic energy. It could be seen that the energy ranges of the three regions vary with the charge (and mass number) of the ion. The results were obtained using the SRIM code [22].

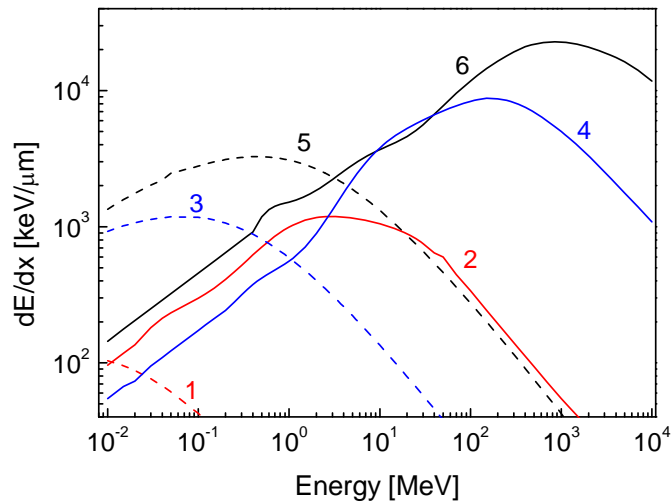

Figure 1: Nuclear (dashed lines) and electronic (continuous lines) stopping powers of C (curves 1 and 2), Ge (3 and 4) and Pb ions (5 and 6) in Si

A detailed discussion on this subject, as well as on the development of understanding related to phenomena associated with the energy loss could be found, e.g., in Sigmund's paper [23]. Following the classical paper of Lindhard [3], the energy loss of particles in materials could be separated into interactions with the electronic system (ionization) and with the system of the lattice. In fact, the physics is more complicated, due both to the possible excitations of these systems, and to their mutual interaction (electron – phonon). The processes of energy transport in materials are governed by phonon-phonon scattering and by phonon-electron interactions.

Another classification of the phenomena related to the energy loss of projectiles and energy transport in the target regards individual versus collective effects. An important example is related to plasmons, quantized collective oscillations of valence electrons [24]. The full Coulomb interaction could be replaced by a screened interaction, and the long range component which is neglected re-emerges in the form of an additional collective oscillation, this last corresponding to compression waves in the electron gas [25]. The case of collective excitations is important, because for low momentum transfers (i.e. distant collisions according to Bohr's terminology) the projectile interacts simultaneously with many atoms, and the perturbed atoms will react with projectile's field affecting its interaction with any other atom of the medium. This way, energy is directly transferred far from the trajectory of the incoming particle. The plasmon is characterised by a significant amount of energy and a little momentum, and its creation is possible in a binary process too, in accord with constraints of energy and momentum conservation.

### 3. Model for transient processes

The energy loss of the incoming particle is eventually imparted between the electronic and lattice subsystems of the target, but, to obtain a simplification of the problem, one of the processes of energy transfer is usually considered as being dominant. It is possible to find a minimal size of spatial region where the primary interactions have place $\Delta x_{min}(\beta) \geq \sqrt{\beta c \hbar / (dE(\beta)/dx)}$ and thus the minimal time for interaction is: $\Delta t_{min}(\beta) \geq \sqrt{\hbar/(\beta c)/(dE(\beta)/dx)}$. In crystalline materials, impact parameter and lattice distance impose supplementary conditions for minimal distances.

Recently, the authors developed a spike model [26] with two heat sources, one from the ionizing energy loss, the other from nuclear energy loss, suitable for ions and energy ranges where nuclear and electronic energy losses are comparable. This model was a generalization of the idea of Izui, Chadderton and Toulemonde [27 – 29] for high ionization energy loss.

In the present paper, the spikes are described mathematically using the homogeneous heat equation:

$$\rho c_p \frac{\partial T(r,t)}{\partial t} = \nabla \cdot (k \nabla T(r,t)) \tag{1}$$

with T(r,0)=$T_0$, the initial temperature of the target, and with the normalisation:

$$\frac{3}{2} N k_B \int (T - T_0) dV = E \qquad (2)$$

where ρ is the mass density of the target, $N$ is the atomic concentration, $c_p$ its specific heat at constant pressure, $k$ the thermal conductivity, $k_B$ the Boltzmann constant, and $E$ the energy transferred in the interaction. $c_p$ and $k$ generally depend on temperature, as well as their ratio the diffusivity $d = k/(\rho c_p)$. This is correct if the medium could be treated as a continuum, i.e. when the diffusion length is large compared to atomic distances. In fact, the treatment is at least suitable for the qualitative description of the thermal spike.

The simplest case has been treated by Seitz and Koehler [13] which made the assumption that thermal parameters characterising the material are temperature independent, and consequently solved the diffusion equation. Vineyard [30] considered that both specific heat and thermal conductivity of the medium have power law dependences on temperature of the type:

$$c_p(T) = c_0 T^{n-1}$$
$$k(T) = k_0 T^{n-1}$$

with $n$ a positive number, the same for both physical quantities: the result is that the diffusivity is also temperature independent.

The problem is that the diffusivity of the lattice is temperature dependent for all types of materials, crystalline or amorphous, metals, insulators and semiconductors [31]. Experimental data suggest a power law dependence of the diffusivity on temperature.

Lattice thermal conductivity is due primarily to acoustic phonons and could be derived considering phonon-phonon collisions. For example, in crystals at very low temperatures $k(T)$ has a $T^3$ temperature dependence, reaches a maximum and than decreases with T first due to phonon scattering via normal ($N$) processes, than due to Umklapp ($U$) processes. The ranges of validity of the $T^3$ law, and of 1/T dependence, as well as the region where the maximum is situated depend all on the Debye temperature $\theta_D$. [32, 33]. In what regards the lattice specific heat, a reasonable agreement of experimental data with the predictions of the Debye model exists: a $T^3$ dependence at cryogenic temperatures and a nearly constant specific heat for $T >> \theta_D$. Consequently, theoretical considerations based on Debye theory lead to the conclusion that in crystals the diffusivity is nearly temperature independent at very low temperatures, and has a *1/T* dependence at high temperatures.

In the present model the time and space profile of the thermal spike produced by the nuclear energy loss (as source of transient processes) is derived analytically as solution of eq. (1) considering that the diffusivity has a power low dependence on temperature. We found that the differential equation could be solved analytically using a Boltzmann transformation, dimensional analysis or otherwise.

### 3.a Cylindrical symmetry

In this approximation, the energy is assumed to be deposited along the ion track uniformly, with constant energy loss $dE/dx$. The 'cylindrical' spike model was first developed in Refs. [30, 34].

In this approximation the problem to be solved for the function $T(r,t)$ is:

$$\rho c_p \frac{\partial T(r,t)}{\partial t} = \frac{1}{r}\frac{\partial}{\partial r}\left(kr\frac{\partial T(r,t)}{\partial r}\right) \quad (3)$$

with the conditions:

$$T(r,0) = T_0$$
$$\frac{3}{2} N k_B \int_0^\infty (T(r,t) - T_0) 2\pi r\, dr = \frac{dE}{dx} \quad (4)$$

If the diffusivity $d = k/(\rho c_p)$ has a power law dependence on temperature:

$$d = d_0 T^n, \quad (5)$$

than the temperature dependence on time and radius for <u>positive n,</u> is obtained as:

$$T(r,t) = \begin{cases} T_0 + A\left(\dfrac{dE}{dx}\right)^{\frac{1}{n+1}} t^{-\frac{1}{n+1}} \left(1 - \dfrac{r^2}{r_0^2}\right)^{\frac{1}{n}} & r > r_0 \\ 0 & \text{otherwise} \end{cases} \quad (6)$$

Here $r_0$ is the distance where the front of heat propagation arrived, and could be written as:

$$r_0(t) = \sqrt{\frac{n+1}{n}} \cdot 2^{\frac{n+2}{2(n+1)}} \left(\frac{1}{3\pi N k_B}\right)^{\frac{n}{2(n+1)}} d_0^{\frac{1}{2n+1}} \cdot \left(\frac{dE}{dx}\right)^{\frac{n}{2(n+1)}} t^{\frac{1}{2(n+1)}} \quad (7)$$

with $A$ depending on $n$, on $d_0$ and on physical constants.

In fact, minimum values for distance and time in the spike evolution might be introduced in accordance with the conditions imposed by the Heisenberg principle.

For $n = 0$, the 2D solution of the classical diffusion equation is obtained, while for <u>negative n, but larger than −1,</u> with $m = -n$, the solution is:

$$T(r,t) = T_0 + A\left(\frac{dE}{dx}\right)^{\frac{1}{1-m}} t^{-\frac{1}{1-m}} \left(r_0^2 + r^2\right)^{-\frac{1}{m}} \quad (8)$$

For $n < -1$, the integral in eq. (4) diverges, and an analytical solution is no more possible.

Due to the existence of an analytical solution, important conclusions could be drawn related to the dependence of the temperature on distance, on time and on energy, corresponding to different values of $n$.

For positive $n$ the heated zone finishes at a distance given by eq. (7). For $n \leq 0$, the temperature asymptotically approaches the temperature of the medium at large distances.

The impossibility to fulfil eq. (4) for $n < -1$ represents a limitation of the present model, and must be understood as being due to the neglect of some phenomena. Cattaneo and Vernotte [35, 36] introduced a supplementary term in the heat equation, containing the second derivative of the temperature in respect to time multiplied by a time constant having the signification of time needed to accumulate energy for significant heat transfer between structural elements. This way the parabolic eq. (1) is transformed into a hyperbolic one, and the paradox of instantaneous heat propagation is also eliminated. A more evolved model is the dual phase lag model [37], which leads to the introduction of another characteristic time in the equation of heat conduction. This way, the classical theory of diffusion (macroscopic in space and time, eq. (1)), is transformed, considering phonon-scattering and phonon-phonon interactions, to a microscopic one, both in time and space.

### 3b Spherical symmetry

The problem could be considered as having spherical symmetry in the case of pointlike interactions well separated in time and distance from the preceding and following ones, i.e. for isolated nuclear interactions.

In this case the temperature is the solution of the homogeneous heat equation, with spherical symmetry:

$$\rho c_p \frac{\partial T(r,t)}{\partial t} = \frac{1}{r^2}\frac{\partial}{\partial r}\left(kr^2 \frac{\partial T(r,t)}{\partial r}\right) \tag{9}$$

The energy $E$ is transmitted in the interaction in the form of an increase in temperature at the origin, which propagates along the radius.

$$\begin{aligned} T(r,0) &= T_0 \\ \frac{3}{2} Nk_B \int_0^\infty (T(r,t)-T_0)4\pi r^2 dr &= E \end{aligned} \tag{10}$$

For *positive n*, the temperature could be written as:

$$T(r,t) = T_0 + BE^{\frac{2}{3n+2}} t^{-\frac{3}{3n+2}} \left(1 - \frac{r^2}{r_0^2}\right)^{\frac{1}{n}} \tag{11}$$

for $r < r_0$, and zero otherwise. $B$ is a constant depending on the exponent $n$, on $d_0$ on the atomic concentration $N$ in the target and on physical constants.

Thus, $r_0$, the distance where the front of heat propagation arrives at time $t$, depends on the energy imparted in the interaction and on time as:

$$r_0(t) \sim E^{\frac{n}{3n+2}} \cdot t^{\frac{1}{3n+2}} \tag{12}$$

For *n=0*, the classical solution of the diffusion equation is retrieved, while for $-2/3 < n < 0$, eq (3) still has solution, of the form:

$$T(r,t) = T_0 + CE^{\frac{2}{2-3m}} t^{-\frac{3}{2-3m}} \left(1 + \frac{r^2}{r_0^2}\right)^{-\frac{1}{m}} \tag{14}$$

where $m = -n$

We note also that for positive *n*, the temperature profile develops in time as $<r^2> \sim t^{\frac{1}{n+1}}$ in the case of cylindrical symmetry, and as $<r^2> \sim t^{\frac{2}{3n+2}}$ in the case of spherical one.

## 4. Discussion of the model results and possible applications

The model is a good approximation for the thermal spike if two types of constraints are simultaneously fulfilled: one is related to particles and materials where the electronic energy loss could be neglected in respect to the nuclear one, and the other refers to materials where the exponent of the diffusivity makes the cylindrical or spherical approximations applicable.

While the specific heat of all materials is a monotonically increasing function of temperature with a plateau at high temperature, the thermal conductivity is in general a function with a maximum, and this way the condition for the exponent of diffusivity $n > -1$ or even $n > -2/3$ is not satisfied in all situations. From this point of view, the model is directly applicable to insulators as glasses, to amorphous semiconductors (e.g. Ge), and to skutterudite compounds [31]. The model is also applicable to some polycrystalline materials, particularly to Si [31, 38]. Also, in the study of sputtering from spikes, in Ref. [39] the value 1/2 for *n* has been used. For single crystal semiconductors, such as Si or Ge, the direct applicability of the model is limited to some temperature ranges. The same is true for diamond [38, 40]. For Si and Ge in thin layers of single- or polycrystals, the exponent of the diffusivity makes the model applicable up 400 K, starting from low temperatures [41, 42].

A difficulty related to spike calculations is the lack of experimental data on temperature dependence of material characteristics on the same sample, taking into account the variability of material parameters.

a. A very interesting material where the present model could be applied is diamond. It has very special thermal properties – a very high thermal conductivity (the highest thermal conductivity at RT) [43]. Combined with some other extreme physical properties, diamond thin films are now considered as the 21st-century material [44]. Thermal conductivity and specific heat of diamond (single crystal and thin films obtained by chemical vapor deposition) depend on impurity content and in the last case also on the microstructure. Extensive characterizations of the temperature dependence of both heat capacity [45] and thermal diffusivity [46, 47] are reported in the literature. There is a relative consensus on the shape of the temperature dependencies of these physical quantities. Using the experimental data from Refs [45, 47] for the heat capacity and thermal conductivity respectively, a value – 0.5 for the exponent *n* of the diffusivity of diamond in the temperature range 30 – 100 K was derived. Considering the thermal spike produced by an ion of Pb in this material, the model of the cylindrical spike is suitable. We present in Figure 2 the spatial and temporal dependencies of the temperature profile produced by an ion of Pb of 2 MeV kinetic energy in diamond, in the plane perpendicular to the trajectory.

Due to the very high thermal conductivity, the energy imparted on the trajectory is quickly spread, both in time and along the radius of the cylinder. Due to the fact that the exponent *n* is negative, the heat spread in the whole space.

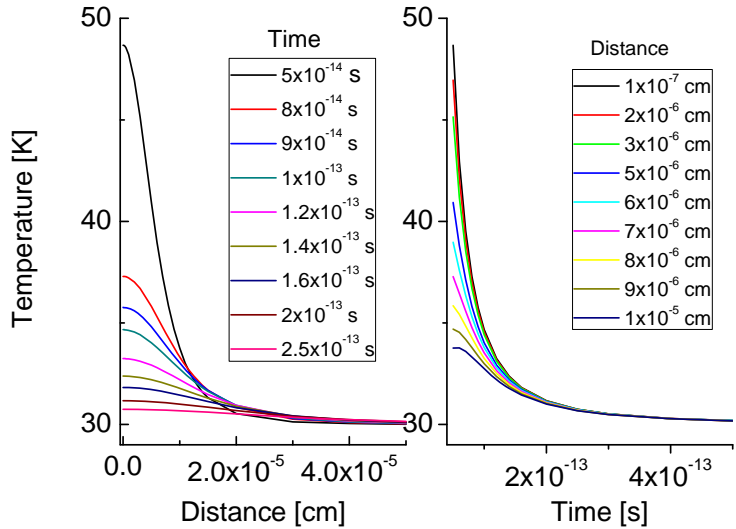

Figure 2: Temperature dependence on distance (a) and on time (b) in the approximation of the cylindrical spike, after the passage of an ion of Pb of 2 MeV in diamond, at 30 K.

b. Amorphous Ge, in the range 200 – 300 K is characterised by an exponent *n* = 0.2 (obtained from the data of heat capacity and thermal conductivity from Refs. [48] and [31] respectively). In the model, irradiating the material with uranium ions of 100 keV kinetic energy, all the energy imparted by the projectile is distributed only into the atomic system. The supplementary consideration of the electronic energy loss and of the coupling between the two subsystems in the solid leads to a more rapid evolution of the spike. This simplified model gives a good approximation of the development of the temperature distribution, especially in space.

c. The spike model could be used in the analysis of the data related to the direct detection of the dark matter in astrophysics experiments. There exists now a major interest in the search for hypothetical WIMPs (Weakly Interacting Massive Particle) which are attractive candidates to account for the cold dark matter in our Universe, one of the detection techniques being the a combination between calorimetric (bolometric) and ionisation measurement method using silicon and/or germanium crystals as cryogenic detectors.

Usually the masses for WIMPs are supposed to be in the range between 10 GeV and 1 TeV; they interact only weakly with ordinary matter via elastic scattering, imparting energy to nuclear recoils, in the range between 10 and 100 keV, and do not directly ionize the medium [49, 50]. Due to the weak coupling and to the low imparted energy, WIMP signals are however rare and elusive events. Recently [51], two candidates for WIMP interaction were communicated.

Practically, most probably one WIMP produces only one interaction in the detector, thus the spherical spike model is justified if the recoil nucleus is not displayed from the lattice. In Figure 3 we present the space and time development of the spike considering that the energy is imparted spherically to Si, at 25 mK. The theory predicts a nearly temperature independent diffusivity for Si and Ge at cryogenic temperatures [31, 52], so that $n = 0$ was used.

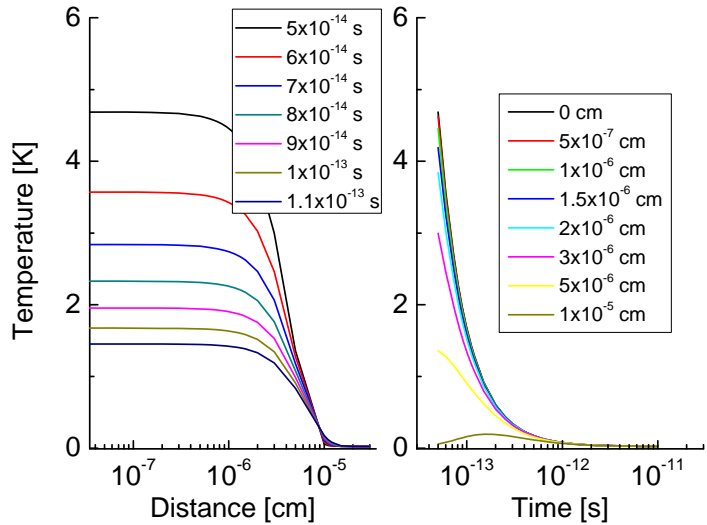

Figure 3. Spherical spike produced by a WIMP which imparts an energy of 15 KeV in Si at 25 mK

If the energy is transmitted to a nucleus, the recoil of about 15 keV energy interacts with the medium producing a cylindrical thermal spike (and ionization).

In Figure 4 we present comparatively the results obtained from the application of the present analytical model and from a previous model of authors [26], which considers also the electronic energy loss of the recoil and the coupling between the two subsystems in the material. In the inset of both figures, we represented the dependence of temperature on distance (a) at different moments and of temperature on time at different distances, respectively.

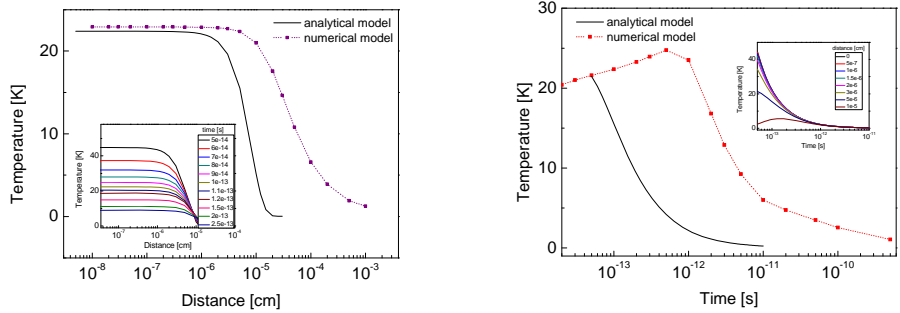

Figure 4. Dependence of temperature on distance at $10^{-13}$ s (a) and on time at $5 \cdot 10^{-6}$ cm (b) in the spike produced by a Si recoil of 15 keV: analytical model of the present work and numerical model from Ref [26]. In the insets the development of the spike is presented.

It could be seen that this simple model predicts a more limited region of high temperatures both in time and in space in respect to the numerical model.

## 4. Summary and conclusions

The mechanisms by which heavy particles lose energy in media are discrete processes due to quantum phenomena, are complex and incompletely understood. An essential aspect is represented by the transient phenomena after the deposition of energy in a primary act of interaction of the incident projectile.

In the present paper, the authors developed an analytical model for the nuclear thermal spike, for power law dependence of the diffusivity on temperature, for cylindrical and spherical symmetries of the problem. The analytical model neglects the electronic contribution to the energy transfer, and also electron-phonon coupling.

The cylindrical model is valid for uniform energy deposition along particle's path, with constant stopping power, corresponding to low energy ions and near the end of their range, in targets and temperature ranges where the exponent of the diffusivity on temperature is greater than $-1$.

The spherical model is valid for the case of rare interactions, well separated in time and distance,

The class of materials to which the model is applicable covers insulators as glasses and amorphous materials in the whole range of temperatures. For usual semiconductors, such as Si or Ge, or for diamond, the model is directly applicable in limited temperature ranges, at low and high temperatures. In thin layers of single- or polycrystals of Si and Ge the exponent of the diffusivity makes the model applicable up to 400 K

Results of the model are presented for diamond, the material with the highest thermal diffusivity, for amorphous Ge, and also for Si at cryogenic temperatures, with applicability in the detection of WIMPS.

**Acknowledgments**

S.L. would like to thank the Romanian Research Council for financial support, under Project IDEI, 901/2009.

# Bibliography

[33]. A. N. Smith and P. M. Norris, Microscale HeatTransfer, in Heat transfer Handbook vol I, Ed. Adrian Bejan and Allan Kraus, John Wiley & Sons, 2003
[34]. H. Hofsäss, H. Feldermann, R. Merk, M. Sebastian, C. Ronning, Appl. Phys. A 66 (1998)153–181
[35]. C.Cattaneo, Compte Rendus 247, (1958) 431- 433.
[36]. P.Vernotte, Compte Rendus,.252, (1958) 3154-3155.
[37]. D.Y. Tzou, Macro- to Microscale Heat Transfer, the Lagging Behavior.Taylor & Francis, Washington,1997
[38]. E.A. Gutierrez-D, M.J. Deen, C. L. Claeys, Low temperature electronics, physics, devices, circuits, and applications, Academic Press 2001.
[39]. P. Sigmund, C. Claussen, J. Appl. Phys. 52, (1981) 990-993
[40]. H.Pitzer, J. Chem. Phys. 6 (1938) 68 – 70.
[41]. Y. H. Lee,·R. Biswas, C. M. Soukoulis, C. Z. Wang, C. T. Chan, and K. M. Ho, Phys. Rev. 43 (1991) 6573 – 6580
[42]. S. Uma, A. D. McConnell, M. Asheghi, K. Kurabayashi and K. E. Goodson, Int. J. Thermophys 22 (2002) 605-616
[43]. T. R. Anthony, Phil. Trans.: Phys. Sci. & Eng.342 (1993) 245 – 251.
[44]. P.W. May, Phil. Trans. R. Soc. Lond. A **358** (2000) 473–495.
[45]. J. E. Desnoyehs; J. A. Morrison, Philosophical Magazine, 3: (1958) 42 — 48
[46]. Y. Li, R. E. Taylor, and A. Nabi, Int J Thermophys. 14, ( 1993) 285 – 295.
[47]. S Barman and G. P. Srivastava, J. Appl. Phys. **101**, (2007) 123507 – 8.
[48]. H.S. Chen, D. Turnbull, J. Appl. Phys. 40 (1969) 4214 – 4215.
[49]. Kari Enqvist and Kimmo Kainulainen, Physics Letters B  264 (1991) 367-372.
[50]. David G. Cerdeno, Anne M. Green, arXiv:1002.1912
[51]. The CDMS II Collab, Science 327 (2010) 1619 - 1621
[52]. G. S. Kumar, J. W. Vandersande, T. Klitsner, and R. O. Pohl, Phys Rev. B 31 (1985) 2157-2162
13